\documentstyle[preprint,aps]{revtex}
\begin{document}
\draft
\preprint{}
\title
{\bf
NOVEL A-B TYPE OSCILLATIONS IN A 2-D ELECTRON  \\ GAS IN INHOMOGENOUS
MAGNETIC FIELDS}
\author{MICHEL CARREAU$^{a,b}$ and JORGE V. JOS\' E$^{a,c}$\\
$^{a}$ Physics Department,
Northeastern University, Boston, MA 02115 USA\\
$^{b}$ Department of Physics, Boston University, Boston MA 02215 USA \\
$^{c}$ Instituut voor Theoretische Fysica, Princetonplein 5,
Postbus 80006,\\ 3508 TA Utrecht, The Netherlands}
\maketitle
\begin{abstract}


We present results from a quantum and semiclassical
theoretical study of the $\rho_{xy}$ and
$\rho_{xx}$ resistivities of a
high mobility 2-D electron gas in the presence of a dilute
random distribution of tubes with magnetic flux $\Phi$ and radius $R$,
for arbitrary values of $k_f R$ and $F=e\Phi/h$.
We report on novel Aharonov-Bohm  type oscillations in
$\rho_{xy}$ and $\rho_{xx}$, related to degenerate quantum flux
tube resonances, that satisfy the selection rule ${(k_fR)}^2=4F(n+{1\over
2})$, with $n$ an integer. We discuss possible experimental conditions where
these oscillations may be observed.
\end{abstract}

\maketitle
\newpage

Transport in a two-dimensional electron gas (2DEG) in the
presence of weak {\bf inhomogenous}  magnetic fields
has recently been the subject of considerable interest, both experimental
\cite{expt1,loc,expt2}  and theoretical  \cite{theor1}.
This situation has been achieved experimentally by gating the 2DEG system
with a type-II  superconducting layer. Abrikosov vortices are then produced
by applying an external magnetic field  perpendicular to the plane of the
layers. In the {\bf ballistic} transport regime and for low fields, when
the density of vortices is small, clear modifications to the Hall resistance
in the  quantum regime have been measured \cite{expt2}.
Previous theoretical studies of this problem
were restricted to the asymptotic quantum regime
$k_fR\gg 1$ \cite{theor1}
and classical limit  $k_fR\ll 1$ \cite{theor2}.
However, the experiments have covered the
interesting  intermediate $k_fR$ crossover regime, with $F=1/2$.
Here, $k_f$ is the Fermi wave vector and $F=\Phi/\Phi_0$ with,
$\Phi_0=h/2$ the flux quantum.

In this paper we present  a full solution to this problem
for arbitrary $k_fR$ and $F$. Our results identify a series of novel
quantum oscillations in the galvanomagnetic properties of the 2DEG,
that appear to be within the reach of an experimental confirmation.
These oscillations can be seen at intermediate ranges of
$k_fR$ and $F(>1/2)$ values and
are related to the Aharonov-Bohm (AB) effect. The intermediate ranges
of  $k_fR$ are experimentally already achievable (e.g. \cite{expt1,expt2}).
At the end of the paper we discuss two  experimental set-ups that have been
suggested to produce  larger values of $F$.
Here we are interested in the experimental situation considered in
\cite{expt2} where the 2-D electrons  move
ballistically between the flux tubes and the dominant
transport mechanism can be assumed to come from electrons scattering off
individual flux tubes. Under these conditions, as a first approximation,
we can apply the results of linear response theory in the Born approximation
\cite{abrikosov}.  These results are formally the same
as those obtained with the Boltzmann equation  \cite{theor1}.
The weak nonlocal localization limit has already been considered
experimentally and theoretically \cite{loc}.

There are three  important physical contributions to the electronic transport
properties of this system: (i)  For finite $R$ the Lorentz force that leads
to an asymmetry in the scattering process. (ii)
A diffractive force, relevant in the $0<k_fR<1$ regime and first considered
by   Iordanski{\v\i} \cite{iordanski},
that also yields a transversal  contribution to the transport,
and (iii) the standard AB contribution \cite{ab1}.
The Iordanski{\v\i} term in $\rho_{xy}$, which
is not taken into account in the differential cross section,
is due to the scattering of electrons by finite radius flux tubes and
has essentially the same  origin as the AB effect \cite{sonin},
for both are topological in nature and
due to the long range properties of the vector potential.
This means, as we see below, that the contribution from (ii)
to the Hall resistance only depends on the value of $F$  and not
on the specific magnetic flux profile chosen in the analysis.

The modification to the Hall resistivity, $\rho_{xy}$,
due to the inhomogenous field can be represented by a {\it Hall coefficient},
$\alpha$, which is defined by the expression \cite{expt2}
\begin{equation}
\rho_{xy}={\alpha(k_fR,F)} {B\over n_{e}e},
\end{equation}
where $n_e$ is the electron density and $B$ the magnetic field.
In the Born approximation of the Kubo formula  the transport coefficients are
expressed in terms of the scattering cross section $f(\phi)$, with $\phi$ the
electronic scattering angle \cite{abrikosov,theor1}.
Explicit limiting values
of $\alpha$ have been calculated in the extreme quantum
$\alpha(k_fR\ll 1)={1\over2\pi F}\,\sin {(2F\pi)}$ and
semi-classical limits $\alpha(k_fR \gg 1)=1$ \cite{theor1,theor2}.
We use these two results as constraints  to be satisfied in our calculations.
Previous studies were restricted to these two limits because of
mathematical difficulties in the evaluation of the scattering amplitude
$f(\phi)$ in the whole $k_fR$ and $F$ ranges.
These difficulties were identified by {Khaetski\v\i } \cite{theor1} and are
essentially related to the singularity in the AB and {Iordanski\v\i}
scattering in the forward direction. In the AB case
 $f_{AB}(\phi\sim 0)\sim 1/{\sin (\phi/2)}$, which would lead
to an infinite $\alpha$  \cite{theor2,iordanski}.

Below we present the results of an explicit evaluation of $f(\phi)$ in the
whole range of $k_fR$ and $F$ values. More importantly,
we use these results to calculate $\alpha$ and
the  magnetoresistance $\rho_{xx}$ in the extended parameter range.
Since the calculational problems arise in the forward
scattering region we consider the  regularized scattering amplitude
\begin{equation}
f_\epsilon(\phi)={e^{-i\pi/4} \over \sqrt{2\pi k_f}}
\sum_{m=-\infty}^{\infty}
\,e^{im\phi}\,e^{-|m|\epsilon}\,[e^{2i\delta_m}-1],
\end{equation}
with $\epsilon$ the regularization parameter, which is taken to
zero at the end of the calculations. Here $\delta_m$ is the phase
shift  associated with the $m^{\hbox{th}}$ partial wave
and can be written as
$\delta_m=\delta^{AB}_m-\tilde\delta_m$.
The $\delta^{AB}_m={\pi\over 2}(|m|-|m+F|)$
  accounts for the AB phase shift,  and $ {\,\tilde\delta_m}=
\tan ^{-1}({{b_m}/{ a_m}})$ for the remaining contribution to the scattering.
The coefficients $a_m$ and $b_m$ are obtained from the asymptotic wave
function solution to the Schr$\ddot o$dinger equation
$\psi(r\to\infty,\theta)\approx \sum_{m=-\infty}^{\infty} \left[
a_m\, J_\nu(k_fr)+b_m\, N_\nu(k_fr)\right]\,e^{im\theta},$
which has the required form for incoming plane waves and outgoing circular
waves. Here $J_\nu(k_fr)$ and $N_\nu(k_fr)$ are the Bessel and
Neumann functions of order $\nu=\vert m+F\vert$
(see Ref. \cite{ab2} for more details).

Including the contributions (i-iii) we can then write the Hall coefficient as,
\begin{equation}
\alpha ={k_f\over 2\pi F}\lim_{\epsilon\to 0}
        \int_{0}^{2\pi} \sin {\phi}\,
        |f_\epsilon(\phi)|^2\, d\phi+ {1\over \pi F}
        \,\sin {(F\pi)}.
\end{equation}
The first term is the regularized Boltzmann contribution while we wrote
the second topological term following {Iordanski\v\i}\cite{iordanski}.
An important property of this expression is that it fully reduces to the
extreme quantum and semiclassical results mentioned above.
In this expression, as long as $\epsilon$ is finite, there is no singularity
in $f_\epsilon(\phi)$ and we can perform the integral. After
evaluating this integral using Eq.(2),
and from the general fact that \break $[\delta_{m+1}
-\delta_{m}]\to 0$ in the limit $\vert m\vert\to \infty$,
we get the {\bf finite} result
\begin{equation}
\alpha  =-{1\over 2\pi F}\,\sum_{m=-\infty}^{+\infty}\,
         \sin {[2(\delta_{m+1}-\delta_{m})]}.
\end{equation}
This equation is one of the main results of this paper, for it provides an
algorithm to calculate $\alpha$ for arbitrary values for the parameters
$k_fR$ and $F$. The number of terms needed in the sum will depend on the
parameter range considered.

The corresponding linear response theory result for the magnetoresistance
$\Delta {\rho_{xx}}$ is
\begin{equation}
{{\Delta {\rho_{xx}(F)}}\over {\rho _{xx}(0)}}={{{\tau_i}\over\tau}-1} =N_F
\ell _i\,\lim_{\epsilon\to 0}
        \int_{0}^{2\pi} (1-\cos \phi )\,
         \vert f_{\epsilon}(\phi)\vert^2\,
d\phi
\end{equation}
\begin{equation}
               ={2\over k_f}N_F \ell _i\,\,\sum_{m=-\infty}^{+\infty}\,
                 \sin ^2{(\delta_{m+1}-\delta_{m})}
\end{equation}
where $\tau$ is the time between electronic flux tube collisions, within the
Kubo-Born approximation, and $N_F$ the concentration of magnetic flux tubes.
When the magnetic field is zero
the transport scattering time is ${\tau_i}={\ell _i}/{v_f}$,
with $v_f$ the Fermi velocity
and $\ell _i$ the impurity scattering elastic mean free path,
which is assumed to be much larger than $ 2R$.

We now proceed to present our results from the direct evaluation of Eqs.(4)
and (6). Later in this paper we present a physical
explanation of these results using a semi-classical analysis.
In plotting these results we have used the
typical experimental parameter values given in the captions.
In Fig.1(a) we show $\alpha$ as a function of $k_fR$
for different values of $F$. The curve $F=1/2$ corresponds to an extended
range of  Fig. 3 in Ref. \cite{expt2}.
We note that for values of $F\leq 1$,
$\alpha$ is a monotonic function of $k_fR$. Notice, however, that for $F=3/4$,
$\alpha$ can become negative for small values of $k_fR$, which comes from
our careful treatment of the extreme quantum  region.
For $F\geq 2$ we see clear oscillations in the $\alpha $ vs $k_fR$ curves
\cite{preprint}.
For $F=10$, for example, we can clearly identify sharp oscillations
of $\alpha $ vs $k_fR$, although their absolute value is smaller.
The number of  oscillations
as a function of $k_fR$ is equal to the integer part of $F$, $[F]$.
For $F=10$, there are
ten oscillations (five of them not shown occur for $k_fR>15$).
Moreover, for small $k_fR$ there are  narrow oscillations superimposed on
the first few oscillations.
In Fig.1(b) we show the Hall resistivity as a function of $F$
 and $k_fR=10$.
For small values of $F$ we see the classical linear behavior of
$\rho _{xy}$ up to a maximum value, after which it decreases as
$F$ increases. We note that the quantum curve, obtained using
Eq.(4),  decreases non monotonically as $F$ increases and even becomes negative
for values of $F\sim 50$. Finally, in Figs.2(a) and
2(b) we show the corresponding results for
${{\Delta {\rho_{xx}(F)}}/{\rho _{xx}(0)}}$
as a function of both $k_fR$ and $F$.
In Fig. 2(a) for $F\leq 1$ we see that
${{\Delta {\rho_{xx}(F)}}/{\rho _{xx}(0)}}$
is a monotonic decreasing function of $k_fR$, while for
larger values of $F$ the magnetoresistance becomes an oscillatory
function of $k_fR$. In Fig.2(b) we note the sharp resonances
that occur exactly at the same values of the minima in $\rho_{xy}$.

We now provide a physical interpretation of these results
in the semi-classical limit. To understand the semi-classical analysis, we
start by discussing the classical problem \cite{annals}.
We note that the energy, $E={1\over 2}m^*v_f^2$, and the total (particle+field)
angular momentum, $J=m^*v_fb-e\Phi$, are constants of the motion.
Here $b$ is the impact parameter and $m^*$ the electron's effective mass.
The impact parameter is defined as $b=y(t\to -\infty)$, where $y$ is along the
direction of the current. The {\bf classical} Hall coefficient is
characterized by the important parameter
$\beta\equiv\omega_cT={{e\Phi}/({2\pi m^* v_fR}})=F/k_fR$.
Different scattering events have different total
angular momenta and different $\beta$ parameters. When $\beta\ll 1$,
the electron trajectories are only slightly affected by the magnetic flux tube.
As $\beta$ increases the Lorentz force becomes important
until a critical $\beta_c$, above which trapped orbits can exist.
The particular quantitative value of $\beta_c$
 depends upon the particular  flux tube profile. For our flux-tube
model  $\beta_c=1/2$, and the trapped circular
orbits have radius
$r_o=R/2\beta$. Both the quantum and classical scattering problems can be
separated into angular and radial components. The radial component
of the classical equation is, as usual, a one-dimensional problem with
effective potential,
$V_{eff}(r,b)= {\left [J+\beta\int^r_0 r'B(r')\,dr'\right ]^2
/{2r^2} }.$
Here the magnetic field of the flux tube is $\vec B(r)=B(r){\vec z}$,
and we have rescaled energies by
$m^*v_f^2$, the angular momentum by
$2\pi mv_fR=h k_fR$, and distances by $R$.
In these units $J=b-\beta$ and the flux tube radius is equal to 1.
In the inset of Fig.1(b) we show curves for
$V_{eff}(r,b)$ for different values of $J$.
We see that electrons with different $J$'s,
or impact parameters,  experience  different effective potentials.
As the total angular momentum decreases, $V_{eff}(r,b)$
develops a potential barrier with height $[J+\beta]^2/2$, which
decreases rapidly.  We can show that for $\beta\ge\beta_c=1/2$
there is a range of $J$'s for which there can be trapped circular
orbits inside the flux tube from $J_1={1\over {4\beta}}$
decreasing to $J_2=1-\beta$.
As $J$ decreases from $J_1$ to $J_2$, the center of the electronic circular
orbit shifts from $r_1=0$ to $r_2=1-r_0=1-{1\over{2\beta}}$.
This range of possible total angular momenta is such that the circular orbit
stays completely within the flux tube.
Classically, these circular orbits can not be reached
by a scattering process. However,
quantum mechanically the scattering electron can tunnel
through this potential barrier and form a quasi-bound state inside the flux
tube for a finite time, and then escape again.
In the classical calculation of the Hall resistivity
and magnetoresistivity shown by dashed lines in the figures,
we computed the classical differential cross-section which
was used in Eqs.(1), (3) and (6), in place of $\vert f(\phi)\vert^2$.

In the semi-classical analysis we associate a classical circular
orbit to each quasi-bound state. Using the standard
Bohr-Sommerfeld quantization condition it is easy to  show that the
quasi-bound states are degenerate and occur at quantized values  of
the  energy, $E_n/\hbar\omega_c=(k_fR)^2/4F=n+1/2$,
with $n$ an integer. This result is
the analog of the Landau level condition in a homogeneous magnetic field.
The factor $n+1/2$ gives the total number of flux quanta enclosed by the
circular orbit. Since the  quanta of flux in the tube is equal to $F$, the
quantum number $n$ ranges from 0 to $[F]-1$.
Moreover, the quantized total angular momentum, $J_m=m\hbar$, leads to
a degeneracy of the $n$ levels. This degeneracy is
equal to the total number of quantized circular orbits
which we can put inside the flux tube.
{}From the range of classically allowed circular orbits mentioned above,
we deduce that the allowed $m$ values start at
$m_1=[(k_fR)^2/4F]=n$ and decrease down to $m_2=[k_fR-F]+\delta$,
with $\delta=1$ if $[k_fR-F]>0$ and $\delta=0$ if $[k_fR-F]\le 0$.
Therefore, we conclude
that the degeneracy is equal to $m_1-m_2+1=n+1+[F-k_fR]-\delta$.
In the figures, the arrows indicate the position of the principal
quantum number $n$ calculated
using the selection rule $(k_fR)^2 = 4F(n+1/2)$. We observe
that they are remarkably well aligned with some maxima and minima of
$\rho_{xx}$ and $\rho_{xy}$. Furthermore, we have numerically determined
that each resonance in $\rho_{xx}$ and $\rho_{xy}$ occurs
at a preferential angular momentum $J_m=m\hbar$ for some $m$.  We arrived
at this conclusion by evaluating the time delay
$t^m_D(k_fR,F)=2\hbar({\partial\delta_m}/{\partial E})
=({2R/v_f})[{\partial\delta_m}/{\partial (k_fR})]$ and we found
that as a function of $m$, $t^m_D$ becomes sharply peaked for one particular
value of $m$ each time the pair $(k_fR,F)$ corresponds to a resonance in
the transport coefficients.  In the figures  we have indicated the
values of $m$ for the resolved resonances.
The minima and maxima in $\rho_{xy}$ correspond
to quasi-bound states due to the tunneling of the
electron into the flux tube.  Semi-classically, as $m$ decreases the resonances
correspond to rotationally asymmetric orbits leading to larger $\rho _{xy}$,
as observed in Fig.1(b). Note that the number of resonances observed for a
particular quasi-bound state level should be equal to the degeneracies of
this level.  However, near $m_1$, the amplitude of the resonances is
suppressed due to the exponentially small tunneling probability
through the potential barrier $(=[J+\beta]^2/2$).  For example note that
$\rho_{xy}=0$ at $F=50$ in Fig.1(b). On the other hand, for
each quasi-bound state level, the observed resonances with the
smallest m (m=3, 0, -7, -51 in Fig.1(b))
correspond precisely to the value $m_2$ derived semi-classically.

We now consider the possible experimental conditions necessary to observe the
galvanomagnetic oscillations described in this paper. The variation of
$k_fR$ in the ranges of interest has already been achieved
\cite{expt1,expt2}. New techniques need to be developed to produce larger
values of $F$ inside the flux tubes. We discuss a couple of possibilities
that have already been suggested to us. The general idea is
to have the usual Hall bar shown schematically in the inset of Fig.2(a),
with the inhomogenous magnetic  field produced by a dilute distribution
of magnetic flux tubes of strength $F$.
One possible way to get larger values of $F$ experimentally is by depositing
randomly located submicron size  {\it superconducting dots} or {\it pillars}
on top of the 2DEG, in a manner similar to the way the dot and antidot
systems have been fabricated \cite{geim2,dots}. Alternatively,
one may drill randomly located sumbmicron holes in the superconducting layer
by using electron beam lithography \cite{goldberg}.
In both cases, by
following a magnetic field cooling technique the magnetic flux  may be pinned
inside the dots  thus trapping a large bundle of flux quanta.
As in the antidot systems we do not expect that the oscillations found here
will be significantly affected by temperature or Coulomb effects,
provided the temperatures are sufficiently low and the charging energy effects
are not significant for the superconducting pillars fabricated.

In conclusion, we have presented a detailed analysis of the transport
properties of a 2-D electron gas system in the presence of a dilute gas
of randomly located magnetic flux tubes for arbitrary values of $k_fR$ and $F$.
The main result from our analysis is the presence of novel AB-like
oscillations in the galvanomagnetic properties of the system.
These oscillations are explained in terms of the degenerate resonant levels,
satisfying the selection rule ${(k_fR)}^2=4F(n+{1\over 2})$, due to the
effective trapping potentials produced by the flux tubes.
A more extensive presentation of the results
described here will appear elsewhere \cite{nos}.
\vskip 0.2cm

We thank  C. Rojas, D. Goldberg, A. K. Geim, V. I. Falko,  D. Weiss
and  R. Putnam for very helpful discussions.
This work was supported in part by grants  ONR-N00014-92-J-1666,
NSF-DMR-9211339,
DE-AC02-89ER40509, DE-AC02-76ER03069, the
NSERC of Canada (M.C.) and by RSDF
 Northeastern University  grant.

\newpage

\newpage
\begin{figure}
\caption{
(a) Hall coefficient $\alpha$ as a function of $k_fR$.
Here $R=100$ nm, the density  of flux tubes is
$\hbox{N}_{\hbox{F}}=10^5$ mm$^{-2}$ and the electron concentration is
$n_e=3.98\times 10^{10}$ cm$^{-2}$, with $\Phi = \pi R^2$. Here
$n$ denotes the
flux tube resonances and $m$ the angular momentum degeneracies.
(b) same as in (a) for the Hall resistivity $\rho_{xy}$ as a function of
$F$. The dashed line corresponds to the classical
results. The inset shows the effective potential
$V_{eff}$ for different values of the total angular momentum
(from top to bottom J=1.25, 0.5, 0.2, 0, -0.25, -0.75)}
\label{fig1}
\end{figure}
\begin{figure}
\caption{
 ${{\Delta {\rho_{xx}(F)}}/{\rho _{xx}(0)}}$ vs $k_fR$ (a)
and  vs F (b) for the same parameter values  as in Fig.1, with mean free
path $\ell _i=2 \, \, \mu$m. The inset shows the  system considered in
this paper formed by a Hall bar with a dilute random distribution of
perpendicular magnetic flux tubes of strenght $F\Phi_0$. See text for
further details.}
\label{fig2}
\end{figure}

\begin{references}
\bibitem{expt1}
A. K. Ge\v\i m, et al.
Pis'ma Zh. Teor. Fiz. {\bf 50}, 359 (1989).
[JETP Lett. {\b 50}, 389 (1989)].
Y. Avishai and Y. B. Band, Phys. Rev. Lett. {\bf 66}, 1761 (1991).
G. H. Kruithof et al., {\it ibid}
{\bf 67}, 2725 (1991).
A. K. Ge\v\i m, et al.
Solid State Comm. {\bf 10}, 831 (1987).
S. J. Bending and A. K. Ge\v\i m, Phys. Rev. B
{\bf 46}, 14912 (1992).
A. K. Ge\v\i m, et al.
{\it ibid} {\bf B46}, 324 (1992).
\bibitem{loc}
J. Rammer and A.L. Shelankov, Phys. Rev B {\bf 36}, 3135 (1987).
S. J. Bending, K. Von Klitzing, and K. Ploog, Phys. Rev. Lett.
{\bf 65}, 1060 (1990).
\bibitem{expt2}
A. K. Ge\v\i m, S. J. Bending, and I. V. Grigorieva,
Phys. Rev. Lett. {\bf 69}, 2252 (1992) and preprint.
\bibitem{theor1} A. V. Khaetski\v\i, J. Phys.:
Cond. Matter {\bf 3}, 5115 (1991).
 D. A. Kuptsov and M. Yu Moiseev, J. Phys I (France),
1165 (1991).
\bibitem{theor2} S. Olariu and I.I. Popescu, Rev. Mod. Phys. {\bf 57},
339 (1985).
\bibitem{abrikosov}
A. A. Abrikosov, L.\ P.\ Gorkov and I.\ E.\ Dzyaloshinski, {\it
Methods of Quantum Field Theory in Statistical Physics} (Pergamon Press
Ltd. Oxford, 1965)
Hall, New Jersey, 1963), Chap.\ 7, p.p. 323.
\bibitem{iordanski}
S. V. Iordanski\v\i{ }, Zh. Teor. Fyz., {\bf 49}, 225 (1965).
[JETP {\bf 22}, 160 (1966)].
S. V. Iordanski\v\i{ } and Koshelev A. E., {\it ibid}
{\bf 90},  1399  (1986). [{\it ibid} {\bf 63}, 820 (1986)].
\bibitem{ab1} Y. Aharonov, and D. Bohm, Phys. Rev. {\bf 115}, 485 (1959).
\bibitem{sonin} E. B. Sonin, Zh. Teor. Fyz., {\bf 69}, 921 (11975).
[JETP {\bf 42}, 469 (1975)].
\bibitem{ab2}
Y. Aharonov, et al.
Phys. Rev.  {\bf D29}, 2396 (1984).
\bibitem{geim2} A. K. Ge\v\i m, S. J. Bending (private communication).
\bibitem{dots}For example in D. Weiss et al. Phys. Rev. Lett. {\bf 66}, 2790
(1991), {\it ibid} D. Weiss et al. {\bf 70}, 4118 (1993).
and D. Weiss (private communication).
\bibitem{goldberg} D. Goldberg (private communication).
\bibitem{preprint}
The results of this paper were briefly presented in the 1993 March APS
meeting (Bull APS, Vol. {\bf 38}, 401 (1993)). We then received a preprint
from L. Brey and H. A. Fertig in
which they have treated some aspects of the problem discussed here.  However,
they did not observe nor explained the rich structure of the electronic
level degeneracies which are at the core of this paper.
\bibitem{annals} M. Carreau, J. V. Jos\'e, and C. Rojas (preprint)
\bibitem{nos} M. Carreau, and J. V. Jos\'e (in preparation)

\end{references}
\end{document}